\documentclass[12pt]{article}
\usepackage[font=small]{caption}
\usepackage{amssymb, amsmath, amsbsy} 
\usepackage{slashed}
\usepackage{tensor}
\usepackage{graphicx}
\usepackage{float}
\usepackage{dsfont}
\usepackage{color}
\usepackage[affil-it]{authblk}
\usepackage{authblk}
\usepackage{rotating}
\usepackage{booktabs}
\usepackage[left=2cm,right=2cm,top=3cm,bottom=3cm]{geometry}
\usepackage{xcolor}
\usepackage{tensor}
\usepackage{amsmath}
\usepackage{breqn}
\usepackage{ulem}

\author[1,2]{Allan  Alonzo-Artiles \thanks{\texttt{allanartiles@gmail.com}}}
\author[3,4]{Ana Avilez-L\'opez\thanks{\texttt{aavilez@fcfm.buap.mx}}}
\author[3,4]{J. Lorenzo D\'iaz-Cruz\thanks{\texttt{jldiaz@fcfm.buap.mx}}}
\author[2,3]{Bryan O. Larios-L\'opez\thanks{\texttt{bryanlarios@gmail.com}}}

\title{{\bf The Higgs-Graviton Couplings: from Amplitudes to the Action}}
\affil[1]{International Centre for Theoretical Physics (ICTP), Strada Costiera 11, 34151 Trieste, Italy}
\affil[2]{Departamento de Gravitaci\'on, Altas Energ\'ias y Radiaciones
\protect\\Facultad de Ciencias
\protect\\Universidad Nacional Aut\'onoma de Honduras
\protect\\Ciudad Universitaria, Tegucigalpa M.D.C. Honduras}
\affil[3]{Centro Internacional de F\'isica Fundamental, BUAP
\protect\\Ciudad Universitaria, Puebla, Pue. M\'exico} 
\affil[4]{Facultad de Ciencias F\'isico - Matem\'aticas
\protect\\Benem\'erita Universidad Aut\'onoma de Puebla  
\protect \\ Apdo. Postal 1364, C.P. 72000, Puebla, Pue. M\'exico} 

\date{}

\begin{document} 

\maketitle

\begin{abstract}
 In this paper we study  the coupling of  scalar (Higgs) particles ($\phi$)  with gravitons ($h$) and their possible effects. 
 The general form of the 3-point interaction  $\phi(p) h(1)h(2)$ can be derived using the scaling behavior of the spinor variables under the little group; 
 the resulting vertices  exhibit such  simplicity, that some simplifications should be hidden in the expressions obtained from the  
 extended scalar action.  To investigate this, we  study an extended Einstein-Hilbert action that besides the minimal coupling, it also includes 
 terms of the form  $\phi R^2$, $\phi R^{\mu\nu} R_{\mu\nu}$ and $\phi R^{\mu\nu\rho\sigma} R_{\mu\nu\rho\sigma}$,   
 as well as the term  $\epsilon_{\mu\nu \alpha\beta}  \phi_5 R^{\mu\nu}_{\rho\sigma} R^{\alpha\beta\rho\sigma}$ 
 for  the case of a pseudo-scalar  ($\phi_5$). 
 The resulting vertices satisfy KLT-type relations, i.e., they can be written as the square of the coupling of the Higgs with  gluons.
  We find that the amplitude for the  Higgs decay into a pair of gravitons (on-shell) only receives a 
 contribution coming from the square of the Riemann tensor. Similar results are obtained for the 3-body decay 
 $\phi \to h h^* (\to XX)$, with an off-shell graviton ($h^*$) that goes into the final state $XX$. One could expect that these quadratic terms can produce new loop effects, however we find that the new contribution from this non-minimal coupling to the graviton self-energy, also vanishes for on-shell gravitons.

\end{abstract}


\bigskip

\bigskip

\bigskip

\section{Introduction}

Understanding the properties of gravity at the quantum level has been the subject of intensive research for almost a century.
Some progress has been gained in  the infra-red (IR) domain, where one considers quantum gravity (QG)  
on the same footing as the other fundamental interactions, with gravity being mediated by 
a massless spin-2 quantum field, the graviton \cite{Feynman:1963ax, DeWitt1967a,DeWitt1967b,'tHooft:1974bx}. This understanding 
of  weakly-coupled gravity was obtained using standard QFT methods  developed during the 60's and early 70's, such as quantization, renormalization, and  regularization of Yang-Mills theories \cite{Hooft:2016gir}. It was learned then that pure  QG is finite at one-loop, 
but divergences start  appearing at two-loop level in pure gravity, while the inclusion of matter interacting with gravity presents divergences 
already at one-loop, leading to the conclusion that QG is non-renormalizable  \cite{tHooft:1974toh}. 
On the other hand, the progress made on effective field theories, which brought a deeper understanding  of  its philosophy and 
calculation machinery, allowed  to perform loop-calculations even for non-renormalizable  theories, such as chiral 
perturbation theory \cite{Weinberg:2016kyd}.  This in turn paved the way to perform reliable calculations for perturbative quantum gravity (pQG)  at the loop-level  \cite{Donoghue1993,Donoghue1994,Bjerrum-Bohr2002, Bjerrum-Bohr2002a}.

More recently, pQG  has benefited from the progress made on modern helicity amplitude methods 
\cite{ManganoParkeReview, TasiLance, BDKUniarityReview}.
Within the common  approach to QFT, as it is presented in most text-book, the kinematics involves the energy-momenta 4-vectors, 
while the wave  function for spin-1/2 and vector particles involves the Dirac spinors and polarization vectors. 
  However, within the Helicity formalism, one works with a unified framework, where both kinematical variables and the wave functions for 
external particles are dealt using spinor variables, i.e. Weyl spinors.  A variety of results that one could call "perturbative jewels" have been discovered over the years,  which include,  for example, the derivation of simple expressions  for the maximal-helicity violating amplitudes for gluons in Yang-Mills theories (Parke-Taylor),  and also for gravitons from pQG \cite{elvang}. One of the most precious result  is 
 that some amplitudes involving gravitons can be expressed as the square of the corresponding amplitude for gluons, a result that is expressed as: $GR=YM\times YM$,  which summarizes the  relationship between gravity and Yang-Mills amplitudes. This result was derived first at tree-level using results in string theory \cite{KLT},  and is known  as the KLT relations. More recently, it was found that some relations hold at loop-level, which is known as  the double-copy formula or Bern-Carrasco-Johansson (BCJ) relation \cite{BCJ, BCJLoop}.
 Most progress has been made for massless  theories, but more recently the massive case has also been tackled 
\cite{Arkani-Hamed:2017jhn, us_sugra}.

On the other hand, scalar particles play an important role in particle physics and in cosmology, with one case being
experimentally confirmed in the HEP front: the SM Higgs boson detected in 2012 at the LHC; measuring its properties is a 
main target of  current and future colliders. But plenty of other hypothetical scalars have been proposed in the literature, ranging from extra Higgs doublets, axion particles, dilaton, inflatons, to name a few. In this regard, understanding the coupling of scalar particles with gravity seems worth studying. This has been discussed before in many contexts, such as Einstein gravity extended with minimaly-coupled scalars  \cite{Veltman:1975}.
This was discussed in connection with the equivalence theorem and for the identification of divergences in quantum gravity \cite{Dunbar:1995ed}.
The quantum corrections to the Newton potential resulting between gravity interacting scalars have been discussed too
\cite{Hamber1995}. More recently,  the possible corrections to Newton law coming from the squares of the Ricci scalar 
and Ricci  tensor have been discussed \cite{Huber:2019ugz}, where it is found that all amplitudes with two heavy scalars and an arbitrary number of gravitons are not affected by those quadratic interactions. Possible scalar resonances appearing in
graviton scattering has been studied too \cite{Blas:2020och}.

In this paper we are interested in studying a related problem that also involves the coupling of a scalar with  gravitons, including those
quadratic terms in curvature. Namely, we want to study the decay of a scalar into gravitons as a possible effect of extended gravity; 
here, the scalar we are referring to could correspond to any Higgs boson that arises within the SM or some extension, it could even be the inflaton or some other super-heavy scalar particle.   One of our goals is to identify possible differences in the contributions  coming from the quadratic terms involving the Ricci scalar, as well as the Ricci and Riemann tensors. The coupling of the Higgs boson ($\phi$) with the graviton ($h$) will be considered in full generality, for both scalar and pseudo-scalar cases.

 The organization of this paper is the following. First, in Section 2  we consider the scaling behavior of the spinor variables
 under the little group to derive the general form of the 3-point vertex  $\phi(p) h(1)h(2)$, identifying  the allowed helicity 
 combinations of the graviton, namely: $h_i (1), h_i (2) = +2, +2$ or  $-2, -2$.   Then, in Section 3 we  look at the Higgs-graviton action, and derive the Higgs-graviton vertices by expanding the gravitational field around Minkowski space-time, i.e.,
 $g_{\mu\nu}= \eta_{\mu\nu}+ \kappa_G h_{\mu\nu}$. For this, we  consider an effective Lagrangian that  
 contains the Ricci scalar, Ricci tensor, and Riemann tensor, and expand them  up to $O(h^2)$ in the graviton field.
 We discuss then the mass-dimension of the resulting coupling,  the coupling of the scalar particle with the those terms; 
 and also consider the interactions of a pseudoscalar with gravitons. We show that
the resulting vertices satisfy the KLT relations, and can indeed be written as the square of the YM terms.
Then in Section 4 we present our calculation of the Higgs decay width, which could be used to discriminate 
among the different terms of the effective action; however, we find that only the term quadratic in the Riemann tensor
contributes to the amplitude for the decay $\phi \to hh$; this remains valid also for the 3-body decay $\phi \to h h^* (\to XX)$,
 with an off-shell graviton that goes into the final state $XX$. These results suggest that may be in order to find such 
 differences, we may have to consider loop effects, which are discussed in Section 5. In particular, we calculate  the new scalar contribution to the graviton self-energy, i.e. the Feynman diagrams with one scalar and one graviton in the loop, but we find 
 that this also vanishes when the external graviton is on-shell.
Conclusions are presented in section 6.

\section{Constructing the Higgs-Graviton vertices from Amplitudes }

Although the use of Lagrangians,  Feynman diagrams and Feynman rules,  seemed irreplaceable in QFT, this view has changed 
with the surge of the constructible program \cite{Benincasa:2007xk}  (also called the ``bootstrap" \cite{Cheung:2017pzi}), where one can
derive the fundamental interaction  from the general properties of amplitudes and spinors that appear in the S-matrix.

Within this approach, the fundamental kinematical variables are the spinors, 
which can be considered  as the building blocks for the amplitudes of massless particles \cite{elvang}.
The starting point is to express the 4-momenta, $p_{\mu}$, as a $2\times 2$ matrix: 
$ p_{\alpha \dot{\alpha}} = \sigma^{\mu}_ {\alpha \dot{\alpha}}  p_{\mu}$ ($\alpha=1,2$ and $\dot{\alpha}=\dot{1},\dot{2}$), 
using the using the Pauli matrices $\sigma^{\mu}$. 
For massless particles $det(p)=0$, and then the momentum matrix  can be factorized in terms of Weyl spinors, as: 
$ p_{\alpha \dot{\alpha}} = \chi_{\alpha}   \tilde{\chi}_{\dot{\alpha}}$. It can be convenient to use Dirac notation, 
with $\chi_{\alpha} (p_i)  \to | i \rangle$ and $\tilde{\chi}_{\dot{\alpha}} (p_i)  \to [ i | $, both for the momentum and for the 
wave functions associated with external fermions. The inner products of spinors are defined as:
$\langle i j \rangle = \chi^\alpha \chi_\alpha$ and $ [i j ] = \tilde{\chi}_{\dot{\alpha}} \tilde{\chi}^{\dot{\alpha}}$, such that:
$\langle j i \rangle [i j ] = 2 p_i \cdot p_j$.

 One also need to have expressions for the polarization vectors in terms of spinors. These are: 
$\epsilon^+_{\alpha \dot{\alpha}} = -\sqrt{2} \eta_{\alpha} \tilde{\chi}_{\dot{\alpha}} / \langle \eta \chi \rangle$, and
$\epsilon^-_{\alpha \dot{\alpha}} = -\sqrt{2} \chi_{\alpha} \tilde{\eta}_{\dot{\alpha}} / [\tilde{\chi} \tilde{\eta}]$, where $\eta, \tilde{\eta}$ are two reference 
 spinors that are unphysical and thus they should disappear at the end of the calculations.
 Thus, one has a unified framework, where both the kinematics and the dynamics are 
 described with the same type of variables.

Within the constructive program for QFT \cite{Cheung:2017pzi},  one attempts to work out all the $n$-point amplitudes in terms of the basic 3-point amplitudes.
The master formulae for the coupling of three particles with helicities $(h_1,h_2,h_3)$, has been derived by working out the scaling
properties of spinors, which amounts to a consideration of their  transformation properties under the Little group.
Namely, under the little group the momentum does not change, and this is respected provided that the spinors transform as:
 $\chi_i \to t_i \chi_i$, and $\tilde{\chi}_i \to t^{-1}_i \tilde{\chi}_i$, while  $\epsilon^+_{\alpha \dot{\alpha}} \to t^{-2} \epsilon^+_{\alpha \dot{\alpha}}$
and  $\epsilon^-_{\alpha \dot{\alpha}} \to t^2 \epsilon^-_{\alpha \dot{\alpha}}$.
An amplitude, partly composed of external lines which depend upon the momenta of the particles, itself transforms under little group transformations by a change of scale; the amplitude follows the same transformation law as the spinors  themselves and scales homogeneously for each of the particles in question. This scaling behavior produces the master formula for the 3-point amplitude, which is given by:

\begin{equation}
A ( 1^{h_1}, 2^{h_2},3^{h_3} ) = \left\lbrace 
\begin{array}{ll}
 c_{123} \, \langle 12\rangle^{h_3-h_2-h_1}  \langle 13 \rangle^{h_2-h_1-h_3}   \langle  23 \rangle^{h_1-h_2-h_3}, 
 \hspace{2mm} & h_1+h_2+h_3 < 0  \nonumber \\
\tilde{c}_{123} \, [ 12 ]^{h_1+h_2-h_3}  [13 ]^{h_1+h_3-h_2}   [ 23 ]^{h_2+h_3-h_1}, 
\hspace{2mm} & h_1+h_2+h_3 > 0
\end{array}
\right.
\end{equation}

\noindent
where $c_{123}$, $\tilde{c}_{123}$ are constants to be determined, and $\langle ij\rangle$ and $[ij]$ denote the spinor products for particles of momenta $i,j$.

We can apply this formula to derive the coupling of a scalar $\phi$ with a pair of gravitons $h$; for the scalar particle
we have  $h_1=0$,  while for the gravitons one has
$h_{2,3} = \pm 2$. Then, for  the combination of helicities $(h_1,h_2, h_3)=(0,-2,-2)$, one gets:

\begin{equation}
A(1^{0}, 2^{-2},3^{-2}) = c \langle23\rangle^{4}.
\end{equation}

On the other hand,  for the other allowed helicity combination $(h_1,h_2,h_3)= (0,+2,+2)$, the result for the amplitude is:

\begin{equation}
A(1^{0}, 2^{+2},3^{+2}) = \tilde{c} [23]^{4}. 
\end{equation}

In deriving these results we have kept only the terms associated with local terms in the corresponding Lagrangian.
The simplicity of these results is quite remarkable, especially when one compares it with the derivation of graviton interactions based 
on the traditional approach, i.e., from the Einstein-Hilbert action, where the graviton couplings (and wave function) involve a massive 
proliferation of space-time indices.

\bigskip

Furthermore, one can also look at the coupling of the Higgs boson with a pair of gluons, which
are given by:  $A(1^{0}, 2^{-1},3^{-1}) = c_0 \langle23\rangle^{2}$, and  $A (1^{0}, 2^{+1},3^{+1}) = c_0  [23]^{2}$.
We can see that the Higgs-graviton couplings (modulo the coefficients $c,\tilde{c},c_0$) can be expressed as the square of the corresponding Higgs-gluon 
couplings, i.e., $A (\phi hh) = A( \phi gg) * A( \phi gg)$. Thus, the Higgs-graviton couplings obey the KLT-type relations. 

\section{The Higgs-graviton  couplings  from the Action}

We are interested in the coupling of a Higgs particle ($\phi$) (Higgs, dilaton, inflaton, etc) with gravity. 
At the lowest order,  the minimal coupling of gravitons and scalars is described by the following Lagrangian:
\begin{equation}
    {\cal{L}}_{mc} = \sqrt{-g} [ \frac{1}{2} g^{\mu\nu} \partial_\mu \phi \partial_\nu \phi - \frac{m^2}{2} \phi^2] 
\end{equation}
This Lagrangian induces several scalar-graviton vertices, including the 3-point vertex: $\phi\phi h$ and 4-point vertex:$\phi\phi h h$, which
can contribute to the graviton self-energy. Thus, in order to have the 3-point vertex with one scalar and two gravitons $\phi hh$ 
we have to consider a Lagrangian with non-minimal couplings. These couplings could be induced at one-loop if one includes a quartic term 
for the scalar particle, which produces  a cubic scalar self-interaction after SSB, and then the triangle or bubble loop-diagrams can induce the 
interaction $\phi hh$ of our interest. 

In this paper we shall consider an effective Lagrangian that includes the non-minimal scalar-graviton couplings. We shall denote by 
$\phi$ the neutral component of an scalar multiplet   $\Phi$  that transforms under some  gauge symmetry. The effective Lagrangian 
shall contains terms with a quadratic factor, $c_i |\Phi|^2$, which after SSB takes the form:  $c_i |\Phi|^2= c_i (V^2 + 2 V \phi + \phi^2) + ... $, 
where $V$ denotes the v.e.v. of the Higgs field, and only the linear term $ 2 V c_i \phi = \kappa_i \phi $ will be considered. 
We shall consider terms involving the scalar curvature ($R$),  its square ($R^2$),  the square of the Ricci tensor 
($R_{\mu\nu}R^{\mu\nu}$)  and the square of the Riemann tensor $R_{\mu\nu\alpha\beta}R^{\mu\nu\alpha\beta}$. 
Thus, the effective Lagrangian is given as follows:
\begin{equation}
{\cal{L} }_{nm} =  \frac{\sqrt{-g}}{\kappa^2} \left( \kappa_1\phi R + \kappa_2 \phi R^2 + \kappa_3 \phi R_{\mu\nu}R^{\mu\nu} + \kappa_4 \phi 
     R_{\mu\nu\alpha\beta}R^{\mu\nu\alpha\beta} +\dots\right),
\label{eq:LagPureGR-v1}
\end{equation}
where $\kappa^2= 8\pi G_N$, and $G_N$ is Newton's gravitational constant. Alternatively, one could consider a singlet field ($\chi$), 
and then include the linear factor $ c_i \chi $, with $c_i$ denoting a constant.
For pure gravity in 4D,  the Gauss-Bonnet term satisfies:
$R^2 - 4R_{\mu\nu}R^{\mu\nu}+ R_{\mu\nu\alpha\beta}R^{\mu\nu\alpha\beta}= \partial_\mu X^\mu$, i.e., it is a total derivative,
and thus it  suffices to take into account only two of those  terms in the effective lagrangian. 
However, when one couples gravity with a scalar, it is not possible to eliminate its effects in Equation~\eqref{eq:LagPureGR-v1}, 
and in principle we can include  all of these terms in the action\cite{Veltman:1975}.

Then, following the usual approach to perturbative quantum gravity, one expands the metric ($g_{\mu\nu}$) upon a Minkowski 
background ($\eta_{\mu\nu}$), with the flcutuations identified as the graviton ($h_{\mu\nu}$), and substitute this expansion in
the curvature invariant terms making up the EH action (and beyond). Here, in order to clarify some issues such as the link between 
the classical action, the  Feynman rules for the gravitons and the amplitudes, the  corresponding  
expansion is carried up to second order in $\kappa$.  Namely:
\begin{equation}
g_{\mu\nu} = \eta_{\mu\nu} + \kappa h_{\mu\nu} + \frac{\kappa^2}{2} h_{\mu\alpha}h^{\alpha}_\nu+\dots,
\label{eq:MetricExpansion}
\end{equation}

One can check the dimension of the couplings $\kappa_i$ that appear in hereafter. Notice that the metric is 
dimensionless, while mass-dimension of the graviton field is $[h_{\mu\nu}]=1$, while $[\kappa]=-1$, then $[g_{\mu\nu}]=[\kappa h_{\mu\nu}]=0$. Being the derivative of the metric, the Chirstoffel symbols have mass dimension $[\Gamma]=1$, while the Riemman tensor, the Ricci tensor and the Ricci scalar (being the second derivative of the metric), have mass dimensions
$[R]=2$.  Thus, we have: $[\kappa_1]=-1$, while $[\kappa_{2,3,4}]=-3$. Then, we can redefine $\kappa_i$ in order to introduce a dimensionless coupling, and play with scales (which we take as $V$ and $\kappa > V$). We can play with two possibilities for $\kappa_1$, namely:
$\kappa_1= \lambda_1/V$ or $\kappa_1= \lambda_1 \kappa$. For $\kappa_{j}$ ($j=2,3,4$) we have more options, 
like $\kappa_{j} = \lambda_j \kappa^3$ or $\kappa_{j} = \lambda_j \kappa^2/V$, with the last case being  the coupling of 
the Higgs boson with gravitons, when this coupling arises at one-loop level.

What about the case of a pseudoscalar? Now, the coupling of a pseudoscalar $\phi_5$ with gravitons will arise from the 
term in the action:

\begin{equation}
    {\mathcal{L}}_5 = \frac{1}{4\kappa^2} \kappa_5  \phi_5 \sqrt{-g} \epsilon_{\mu\nu\rho\sigma}R^{\mu\nu}_{\hspace{3mm}\alpha\beta}R^{\rho\sigma\alpha\beta}.
\label{eq:LagPureGR-Ps}    
\end{equation}

\bigskip

We present next the main results of the expansion for each of the terms in the action foe the scalar and pseudo-scalar, 
as well as the resulting interaction vertices written
in momentum space.

\subsection{ $\phi hh$ interaction from $R$ (2nd order) }

We start by considering the expansion of the Ricci scalar $R$. The first term in the expansion  is given by:
\begin{equation}
R^{(1)} = \partial_{\mu}\partial_{\nu}h^{\mu\nu} - \partial^2 h
\end{equation}
where $h= h^{\mu}_{\, \, \mu}$. The superscript in the scalar curvature denotes the order in $\kappa$ that is being
considered. One can see that the first term in the  expansion contains only one graviton,  and thus it does not induce 
the vertex $\phi hh$. But when one considers the next order, we do get two gravitons; namely:

\begin{equation}
    \begin{split}
    (\sqrt{-g}R)^{(2)} &= R^{(2)} + \frac{1}{2}h R^{(1)} \\
    &= \frac{1}{4}h^{\mu\nu}\partial^2h_{\mu\nu} + \frac{1}{2}\partial_{\mu}h^{\mu\nu}\partial_{\beta}h_{\nu}^{\beta} + 
     \frac{1}{2}h\partial_{\mu}\partial_{\nu}h^{\mu\nu} -\frac{1}{4}h\partial^2h \\
    &+ \partial_{\mu}\Big(-\frac{1}{4}h\partial^{\mu}h - \frac{1}{4}h_{\alpha\beta}\partial^{\mu}h^{\alpha\beta} + h^{\mu\beta}\partial_{\beta}h - h^{\mu\beta}\partial_{\nu}h^{\nu}_{\beta}\Big),
\end{split}
\end{equation}

Then, we can find the expression for  the corresponding  vertex  $\phi(p) h(k_1) h(k_2)$ in momentum space. 
We find:
\begin{equation}
\label{eq:HE_vertex_term}
    V^{\{\phi hh\}}_{\mu\nu\rho\sigma}(k,k') = \kappa_1 \left(-\frac{1}{4}{k'}^{2}\eta_{\rho\mu}\eta_{\sigma\nu} + \frac{1}{4}{k'}^2\eta_{\mu\nu}\eta_{\rho\sigma} + k_{\nu}k'_{\sigma}\eta_{\rho\mu} - \frac{1}{2}k'_{\rho}k'_{\sigma}\eta_{\mu\nu}\right).
\end{equation}


\subsection{$\phi hh$ interaction from $R^2$}

The interaction of the Higgs with a pair of gravitons can arise from the term $\sqrt{-g} R^2$. In this case, from the very definition,
one can see that the resulting vertex can be expressed as the product of two similar expressions, namely:

\begin{equation}
    (\sqrt{-g}R^2)^{(2)} = (R^2)^{(2)} = (\partial_{\mu}\partial_{\nu}h^{\mu\nu} - \partial^{2}h)(\partial_{\alpha}\partial_{\beta}h^{\alpha\beta} - \partial^{2}h).
\end{equation} 

The corresponding  vertex $\phi(p) h(k_1) h(k_2)$  in momentum space, is given by:

\begin{equation}
    V^{\{\phi hh\}}_{\mu\nu\rho\sigma}(k,k') =  \kappa_2 \Big(2k_{\mu}k_{\nu}k'_{\rho}k'_{\sigma} -4k^{2}k'_{\rho}k'_{\sigma}\eta_{\mu\nu} + 2k^{2}{k'}^{2}\eta_{\mu\nu}\eta_{\rho\sigma}\Big).
\end{equation}

\subsection{ $\phi hh$ interaction from $R_{\mu\nu}R^{\mu\nu}$}

Similarly, the expansion  of the scalar invariant formed as the contraction of the Ricci tensor
 and its inverse, is given as: 

\begin{equation}
\begin{split}
    (\sqrt{-g} R_{\mu\nu}R^{\mu\nu})^{(2)} &= \frac{1}{4}\partial_{\mu}\partial_{\nu}h\partial^{\mu}\partial^{\nu}h + \frac{1}{4}\partial^2h_{\mu\nu}\partial^2h^{\mu\nu} + \frac{1}{2}\partial_{\mu}\partial_{\nu}h\partial^2h^{\mu\nu} \\
    &- \partial_{\mu}\partial_{\nu}h\partial_{\gamma}\partial^{\nu}h^{\gamma\mu} - \partial^2h_{\mu\nu}\partial_{\gamma}\partial^{\nu}h^{\gamma\mu} \\
    &+ \frac{1}{2}\partial_{\mu}\partial_{\nu}h^{\mu}_{\beta}(\partial_{\alpha}\partial^{\nu}h^{\alpha\beta} + \partial_{\alpha}\partial^{\beta}h^{\alpha\nu}).
\end{split}
\end{equation}

As in the previous case, we can show that this expression can be written as the product of two terms, namely:

\begin{eqnarray}
    ( \sqrt{-g} R_{\mu\nu}R^{\mu\nu} )^{(2)} 
    & = & \Big[ \frac{1}{2} ( h^{\lambda}_{\nu,\mu,\lambda} - h^{\lambda}_{\lambda,\nu,\mu} + h^{\lambda}_{\mu,\nu,\lambda} - 
     h^{ \, \, \lambda}_{\lambda \mu \nu} ) \Big]   \\
    & \times &  \Big[ \frac{1}{2} ( h^{\lambda \nu, \mu}_{\hspace{5mm},\lambda} - h^{\lambda,\nu,\mu}_{\lambda} + 
    h^{\lambda \mu,\nu}_{\hspace{5mm},\lambda} - h^{\mu \nu,\lambda}_{\hspace{5mm},\lambda} ) \Big]
\end{eqnarray}

The corresponding  vertex $\phi(p) h(k_1) h(k_2)$, in momentum space, is given by:

\begin{eqnarray}
    V^{ \{ \phi hh \} }_{\mu\nu\rho\sigma}(k,k') &=&  \kappa_{3} 
    \Big[ \frac{1}{2} (k \cdot k')^2 \eta_{\mu\nu} \eta_{\rho\sigma} + \frac{1}{2}k^2{k'}^{2}\eta_{\rho\nu}\eta_{\sigma\mu} -
              2(k \cdot k')k_{\sigma}k'_{\rho}\eta_{\mu\nu} + (k \cdot k')k_{\nu}{k'}_{\sigma}\eta_{\rho\mu} \\ 
   & &+ k_{\rho} k_{\sigma}{k'}^{2} \eta_{\mu\nu} - 2k_{\nu}k_{\sigma}{k'}^{2}\eta_{\rho\mu} + k_{\mu}k_{\rho}{k'}_{\nu}{k'}_{\sigma}\Big]
\end{eqnarray}

\subsection{$\phi hh$ interaction from $R_{\mu\nu\alpha\beta}R^{\mu\nu\alpha\beta}$}

The expansion of the squared Riemann tensor, gives the following terms:

\begin{equation}
\begin{split}
    (\sqrt{-g} R_{\mu\nu\alpha\beta}R^{\mu\nu\alpha\beta})^{(2)} &= \partial_{\alpha}\partial_{\beta}h_{\mu\nu}\partial^{\alpha}\partial^{\beta}h^{\mu\nu} - 2\partial_{\mu}\partial_{\nu}h_{\alpha\beta}\partial^{\mu}\partial^{\beta}h^{\alpha\nu} \\
    &+ \partial_{\mu}\partial_{\nu}h_{\alpha\beta}\partial^{\alpha}\partial^{\beta}h^{\mu\nu}.
\end{split}
\end{equation}

Again, we can work this expression and show that it can be written as the product of two terms, namely:

\begin{equation}
\begin{split}
    (\sqrt{-g}R_{\mu\nu\alpha\beta}R^{\mu\nu\alpha\beta})^{(2)} 
    &= \Big[\frac{1}{2}(h_{\alpha\nu,\beta,\mu} - h_{\beta\alpha,\nu,\mu} + h_{\beta\mu,\nu,\alpha} - h_{\mu\nu,\beta,\alpha})\Big] \\
    &\times \Big[\frac{1}{2}(h^{\alpha\nu,\beta,\mu} - h^{\beta\alpha,\nu,\mu} + h^{\beta\mu,\nu,\alpha} - h^{\mu\nu,\beta,\alpha})\Big].
\end{split}
\end{equation}

The corresponding  vertex $\phi(p) h(k_1) h(k_2)$ in momentum space,  is given as:

\begin{equation}
    \label{eq11}
    V^{\{\phi hh\}}_{\mu\nu\rho\sigma}(k,k') =  \kappa_4 \Big(2k_{\rho}k_{\sigma}k'_{\mu}k'_{\nu} -4(k \cdot k')k_{\sigma}k'_{\nu}\eta_{\rho\mu} + 2(k \cdot k')(k \cdot k')\eta_{\rho\mu}\eta_{\sigma\nu}\Big).
\end{equation}

\subsection{$\phi_5 hh$ interaction for a pseudoscalar}

The expansion of Equation (6), gives the following result:
\begin{equation}
\begin{split}
    \Big(\frac{\sqrt{-g}}{4} \epsilon_{\mu\nu\rho\sigma}R^{\mu\nu}_{\hspace{3mm}\alpha\beta}R^{\rho\sigma\alpha\beta}\Big)^{(2)} &= -\frac{1}{2}\epsilon_{\mu\rho\sigma\alpha}\partial^{\alpha}\partial_{\beta}h_{\nu}^{\sigma}\partial^{\rho}\partial^{\nu}h^{\beta\mu} + \frac{1}{2}\epsilon_{\mu\rho\sigma\alpha}\partial^{\alpha}\partial_{\nu}h_{\beta}^{\sigma}\partial^{\rho}\partial^{\nu}h^{\beta\mu} \\
    &+ \frac{1}{4}\epsilon_{\nu\rho\sigma\alpha}\partial^{\alpha}\partial^{\sigma}h_{\beta\mu}\partial^{\rho}\partial^{\nu}h^{\beta\mu}.
\end{split}
\end{equation}
\newline

This term leads to the following expression for the  vertex $\phi_5(p) h(k_1) h(k_2)$:

\begin{equation}
    \label{eq13}
    V^{\{\phi_5 hh\}}_{\mu\nu\rho\sigma}(k,k') =  \kappa_5 \Big(\epsilon_{\beta \delta \nu \sigma}k^{\beta}k_{\rho}{k'}^{\delta}{k'}_{\mu} - \epsilon_{\beta\gamma\nu\sigma}(k \cdot k')k^{\beta}{k'}^{\gamma}\eta_{\rho\mu}\Big).
\end{equation}

\section{The Higgs decay into gravitons }

In this section we use the vertices $\phi  hh $ from the previous section to calculate the decay width for the mode
$\phi (p) \to h (k) h (k')$.   Besides obtaining the amplitude for the decay from the extended action, we also want to verify 
that the results agree with the ones derived from the constructible approach (Sect. 2).

We work with the spinor helicity method, which permits to convert the tensorial amplitude into a function of 
spinor products. In order to build the corresponding amplitudes, one has to multiply the interaction vertices $\phi(p) h(k_1) h(k_2)$ 
with the graviton polarization  tensors, which can be expressed  as the product of polarization vectors,  i.e.,  
$\epsilon_{\mu \nu} (k_i) =  \epsilon_{\mu} (k_i) \epsilon_{ \nu} (k_i)$. 
Within the spinor helicity formalism, for a vector particle of momentum $k$,  the polarization vectors  are given by:
$ \epsilon^{\mu}_{+}(k;q)= -\frac{\langle q |\gamma^{\mu}|k]}{\sqrt{2} \langle qk \rangle}$, and
$ \epsilon^{\mu}_{-}(k;q)= -\frac{\langle k  |\gamma^{\mu}| q ]}{\sqrt{2} [ qk ]}$,
where $q$ is a light-like reference momentum ($q^2=0$), it is arbitrary provided that $q \neq k$, 
and it can be chosen in such a way that the calculations are simplified.
We also need to consider the on-shell conditions for the three particles, i.e. $k^2 = {k'}^2= 0$ and $p^2=m^2_{\phi}$, 
 as well as the transversality conditions for the  polarization vectors.

  The first important lesson that we obtain is that the amplitudes for the decay $\phi \to hh$,
  only get contributions from the term containing the square of the Riemann tensor, while the amplitude constructed
 from  the terms  $\phi R$ (2nd order), $\phi R^2$ and $\phi R^{\mu\nu} R_{\mu\nu}$ vanish for on-shell gravitons. On the other hand, 
 we find that  for the pseudo-scalar case,  the term considered in eq. 6 does contribute to the corresponding amplitude.

 \subsection{Helicity amplitudes for $\phi \to hh $ (Scalar)}
 
Thus, we have found that for on-shell gravitons, the only contributions to the amplitude for the decay of the scalar into gravitons 
comes from the squared Riemann tensor, i.e., $ \frac{1}{\kappa^2}\sqrt{-g}\phi R_{\mu\nu\alpha\beta}R^{\mu\nu\alpha\beta}$.
Then, the total amplitude for graviton with polarizations $h_1 = h_2 = +2$, is given by:
\begin{equation}
     \mathcal{A}_{3}(\phi h^{+}h^{+}) = \frac{\kappa_4 }{2}[kk']^4,
\end{equation}

and the total amplitude for the opposite helicity combination is given by: 
\begin{equation}
    \mathcal{A}_{3} (\phi h^{-}h^{-}) = \frac{\kappa_4}{2}\langle kk' \rangle^4.
\end{equation}

Any other helicity combination is zero for all interaction terms; for example, $\mathcal{A}_{3}(\phi h^{+}h^{-}) = 0$. 
These results are summarize in table 1.

\begin{table}[h!]
\centering
 \begin{tabular}{||c|c|c|c|c||} 
 \hline
 Interaction & $\lambda\lambda'$ & $\mathcal{A}_{\lambda\lambda'}$ & $\lambda\lambda'$ & $\mathcal{A}_{\lambda\lambda'}$ \\
 \hline
 $\sqrt{-g}\phi R$ (2nd order) & ++, ++ & 0 & - -, - - & 0\\ 
 \hline
 $\sqrt{-g}\phi R^2$ & ++, ++ & 0 & - -, - - & 0\\ 
 \hline
 $\sqrt{-g}\phi R_{\mu\nu}R^{\mu\nu}$ & ++, ++ & 0 & - -, - - & 0\\ 
 \hline
 $\sqrt{-g}\phi R_{\mu\nu\alpha\beta}R^{\mu\nu\alpha\beta}$ & ++, ++ & $\frac{\kappa_4}{2}[kk']^4$ & - -, - - & $\frac{\kappa_4}{2} \langle kk' \rangle^4$\\
 \hline
\end{tabular}
\caption{Helicity Amplitudes for the 2-body Higgs decay $\phi \to hh$.}
\end{table}

We notice that this result based on the expansion of the extended gravity action, agrees with the 3-point amplitude found in section 2,
which was obtained based on the constructive approach. This is what was expected, but one can not stop wondering about the huge
simplification that was obtained by  bootstrapping  the amplitudes.

 \subsection{Helicity amplitudes for $\phi_5 \to hh $ (Pseudoscalar)}
 
On the other hand,  for the pseudo-scalar case we have studied the graviton expansion of the term
$\sqrt{-g} \phi_5 \epsilon_{\mu\nu\rho\sigma}R^{\mu\nu}_{\hspace{3mm}\alpha\beta}R^{\rho\sigma\alpha\beta}$.
After multiplying the resulting vertex with the polarization tensors, we find a non-vanishing amplitude, and for  
for the helicities $h_1 = h_2 = +2$  it is given by:

\begin{equation}
    \mathcal{A}_{3}^{P.S.}(\phi_{5}h^{+}h^{+}) =   \kappa_5 \frac{[kk']^2}{4\langle kk' \rangle^2}
    \epsilon_{\beta \delta \nu \sigma}k^{\beta}{k'}^{\delta}\langle {k'} |\gamma^{\nu}|k]\langle k|\gamma^{\sigma}|{k'}].
\label{eq:AmplitudePseudo}
\end{equation}

The complex conjugate of Equation~\eqref{eq:AmplitudePseudo} is:
\begin{equation}
    \mathcal{A}_{3}^{P.S.}(\phi_{5}h^{+}h^{+})^{*} = \kappa_5 \frac{\langle k'k \rangle^2}{4[{k'}k]^2}
    \epsilon_{\alpha \lambda \mu \rho}k^{\alpha}{k'}^{\lambda}\langle k|\gamma^{\mu}|{k'}]\langle {k'}|\gamma^{\rho}|k].
\end{equation}

As a result one gets the following expression for the squared amplitude:

\begin{equation}
\label{eq9}
\begin{split}
    |\mathcal{A}_{3}^{P.S.}(\phi_{5}h^{+}h^{+})|^2 &= \mathcal{A}_{3}^{P.S.}(\phi_5 h^{+}h^{+})\mathcal{A}_{3}^{P.S.}(\phi_5 h^{+}h^{+})^{*} \\
    &= \frac{\kappa^2_5}{16}\epsilon_{\beta\delta\nu\sigma}\epsilon_{\alpha\lambda\mu\rho}k^{\beta}{k'}^{\delta}k^{\alpha}{k'}^{\lambda}\langle {k'}|\gamma^{\nu}|k]\langle k|\gamma^{\sigma}|{k'}]\langle k|\gamma^{\mu}|{k'}]\langle {k'}|\gamma^{\rho}|k].
\end{split}
\end{equation}

We simplify the previous expression by using the following identity for the product of the Levi-Civita symbols (obtained with the help of \textit{Mathematica}):
\begin{equation}
\begin{split}
    \label{eq10}
    \epsilon_{\beta\delta\nu\sigma}\epsilon_{\alpha\lambda\mu\rho} &= -\eta_{\alpha\sigma}\eta_{\rho\beta}\eta_{\mu\delta}\eta_{\lambda\nu} + \eta_{\alpha\nu}\eta_{\rho\beta}\eta_{\mu\delta}\eta_{\lambda\sigma} + \eta_{\alpha\sigma}\eta_{\rho\beta}\eta_{\lambda\delta}\eta_{\mu\nu} - \eta_{\alpha\delta}\eta_{\rho\beta}\eta_{\lambda\sigma}\eta_{\mu\nu} - \eta_{\alpha\nu}\eta_{\rho\beta}\eta_{\lambda\delta}\eta_{\mu\sigma} \\
    &+\eta_{\alpha\delta}\eta_{\rho\beta}\eta_{\lambda\nu}\eta_{\mu\sigma} + \eta_{\alpha\sigma}\eta_{\mu\beta}\eta_{\rho\delta}\eta_{\lambda\nu} - \eta_{\alpha\nu}\eta_{\mu\beta}\eta_{\rho\delta}\eta_{\lambda\sigma} - \eta_{\alpha\sigma}\eta_{\lambda\beta}\eta_{\rho\delta}\eta_{\mu\nu} + \eta_{\alpha\beta}\eta_{\rho\delta}\eta_{\lambda\sigma}\eta_{\mu\nu} \\
    &+ \eta_{\alpha\nu}\eta_{\lambda\beta}\eta_{\rho\delta}\eta_{\mu\sigma} - \eta_{\alpha\beta}\eta_{\rho\delta}\eta_{\lambda\nu}\eta_{\mu\sigma} - \eta_{\alpha\sigma}\eta_{\mu\beta}\eta_{\lambda\delta}\eta_{\rho\nu} + \eta_{\alpha\delta}\eta_{\mu\beta}\eta_{\lambda\sigma}\eta_{\rho\nu} + \eta_{\alpha\sigma}\eta_{\lambda\beta}\eta_{\mu\delta}\eta_{\rho\nu} \\
    &-\eta_{\alpha\beta}\eta_{\mu\delta}\eta_{\lambda\sigma}\eta_{\rho\nu} - \eta_{\alpha\delta}\eta_{\lambda\beta}\eta_{\mu\sigma}\eta_{\rho\nu} + \eta_{\alpha\beta}\eta_{\lambda\delta}\eta_{\mu\sigma}\eta_{\rho\nu} + \eta_{\alpha\nu}\eta_{\mu\beta}\eta_{\lambda\delta}\eta_{\rho\sigma} - \eta_{\alpha\delta}\eta_{\mu\beta}\eta_{\lambda\nu}\eta_{\rho\sigma} \\
    &-\eta_{\alpha\nu}\eta_{\lambda\beta}\eta_{\mu\delta}\eta_{\rho\sigma} + \eta_{\alpha\beta}\eta_{\mu\delta}\eta_{\lambda\nu}\eta_{\rho\sigma} + \eta_{\alpha\delta}\eta_{\lambda\beta}\eta_{\mu\nu}\eta_{\rho\sigma} - \eta_{\alpha\beta}\eta_{\lambda\delta}\eta_{\mu\nu}\eta_{\rho\sigma},
\end{split}
\end{equation}

Then, after applying some spinor product identities, we finally get the result:

\begin{equation}
    |\mathcal{A}_{3}^{P.S.}(\phi_{5}h^{+}h^{+})|^2 = \frac{\kappa^2_5}{16}\langle kk' \rangle^4 [kk']^4.
\end{equation}

Thus, we found that the expression for the squared amplitude is actually simpler than for the amplitude itself. 
Then, one would be tempted to write the corresponding amplitude as follows:

\begin{equation}
    \mathcal{A}_{3}^{P.S.}(\phi_{5}h^{+}h^{+}) = \frac{\kappa_5}{4} [kk']^4,
\end{equation}

which would also be of the form derived in section 2. 

\subsection{The decay width for $\phi \to hh $ } 

  In order to obtain  the  decay width for the  two-body mode $\phi \to h \, h$, one has to square the amplitudes (20) and (21) and add them. 
  Thus, the squared amplitude, summed over polarizations ($h_1=\pm2$ and $h_2=\pm2$ ), takes the form:
\begin{equation}
    \langle |\mathcal{M} (\phi \to hh)|^2 \rangle = \sum_{h_i} \kappa^2_4 |A_3(\phi h^{h_i} h^{h_i}) |^2 = \frac{\kappa^2_4}{2} m^8_{\phi}.
    \end{equation}

The expression for the decay width of  the Higgs (in the rest frame) is given as follows:
\begin{equation}
\Gamma(\phi\to hh)=\frac{S | \vec{p} |}{8\pi m_{\phi}^2}\langle|\mathcal{M}(\phi \to hh)|^2\rangle,
\end{equation}
where $|\vec{ p }|$ is the magnitude of the momentum of either outgoing graviton, and $S$ is a factor to account for identical particles 
in the final state; here, $S = 1/2$. The final expression for the decay width is given by:
\begin{equation}
\Gamma(\phi\to hh)=\frac{\kappa_{4}^2m_{\phi}^7}{64\pi}.   
\end{equation}

The decay width for the SM Higgs depends on the value of $\kappa_4$, which arises starting at one loop-level  \cite{Delbourgo:2000nq}; 
for instance, the contribution from the top quark is of the order: 
$\kappa_4 \simeq G_{N}m^2_t / (v m^2_\phi) \simeq m^2_t / ( \kappa^2 v m^2_{\phi})  \simeq 1/ (v \kappa^2)$,
where $m_t$ and $v$ denote the top quark mass and the electroweak vev, respectively. This expression is in agreement 
with our estimate based on dimensional analysis presented in Sect. 2. This formula produces an extremely small decay width, 
of order $10^{-70} GeV$, which is clearly not observable. 
However, for other scalars with heavier masses the decay width may be more sizable. 
Similarly, we can study the corresponding decay width for the pseudo-scalar case, the difference will appear when
one considers the one-loop contribution of the top quark to the value of $\kappa_5$. 


\subsection{The 3-body decay width for $\phi \to hh^* $} 

Part of our interest in studying the Higgs decay into gravitons was the hope that it could help to distinguish the  different contributions 
associated with each  term in  the effective action, but we found that  only the square of the Riemann tensor contributes 
to the decay amplitude for on-shell gravitons. Thus it is valid to ask ourselves whether we could get some contribution from 
those terms in the effective action when we consider some process with one gravitons off-shell, whether a N-body decay width or
 a cross-section. 
Here we look into this question by considering the decay of the Higgs into one real graviton and
one off-shell graviton, which in turn decays into other particles, such as a pair of scalars, fermions or gauge bosons. 
As this coupling of the graviton with matter through the energy-momentum tensor, we can use this in order to discuss 
the process as general as possible. 

The amplitude for the decay $\phi (p) \to h (p_1) + X(p_3) X(p_3)$ is given by:

\begin{equation}
M^{h_i} = \epsilon^{\mu\nu} (p_1) V_{\mu\nu\rho\sigma} (p,p_1, q) \frac{iP^{\rho\sigma\alpha\beta}}{q^2} T_{\alpha\beta}(XX).
 \end{equation}
 where $T_{\alpha\beta}(XX)$ denotes the energy-momentum tensor for X matter. 
The tensor $P^{\rho\sigma\alpha\beta}$ appears in the graviton propagator, which is taken in de Donder or harmonic gauge.
 
 Now, we shall discuss first the contribution from the Ricci scalar (2nd order), as well as the squares of the Ricci Scalar, Ricci tensor and Riemann tensor.
 
\begin{itemize}

\item First, we consider the contribution from the Ricci scalar expanded up to second order in $h$, we get that:
\begin{equation}
    \epsilon^{\mu\nu}(k) _{+}V_{\mu\nu\rho\sigma} = -\frac{\kappa_{1}}{8\langle q p_1 \rangle^2} p^{2}_1 \langle q|\gamma_{\rho}|p_1]\langle q|\gamma_{\sigma}|p_1],
\end{equation}
which vanishes when we take the on-shell condition of the first graviton.

\item For the squared Ricci scalar, we also find that $\epsilon^{\mu\nu}_{+}(p_1)V_{\mu\nu\rho\sigma} = 0$,
so this term would also not contribute in any 3-body decay of the Higgs.

\item Next, we consider  the contribution coming from the squared Ricci tensor; for a graviton helicity $ h = +2$, we find that:

\begin{equation}
    \epsilon^{\mu\nu}_{+}(k)V_{\mu\nu\rho\sigma} = \frac{\kappa_3}{4 \langle qp_1 \rangle^2}p^2_1 q^{2} \langle q|\gamma_{\rho}|p_1]
     \langle q|\gamma_{\sigma}|p_1],
\end{equation}
which vanishes upon taking the on-shell condition for the first graviton, i.e, $p^2_1 = 0$.

\item Finally, we consider the amplitude obtained from squared Riemann tensor. One needs to contract the polarization tensor with the 
corresponding vertex, we find gain that this is the only non vanishing term. Namely, the contraction of the vertex with the tensor 
$P^{\rho\sigma\alpha\beta}$, gives the  result:
\begin{equation}
\epsilon^{\mu\nu} V_{\mu\nu\rho\sigma}  P^{\rho\sigma\alpha\beta} = 
(q\cdot \epsilon^{i})^2 p^{\rho}_1 p^{\sigma}_1 + \frac{1}{4} q\cdot \epsilon^{i} q\cdot p_1
(p^{\rho}_1 \epsilon^{i\sigma} + p^{\sigma}_1 \epsilon^{i\rho}) + (q\cdot p_1)^2  \epsilon^{i\rho} \epsilon^{i\sigma}
\end{equation}

\end{itemize}
 
Thus,  the amplitude  for the 3-body decay $\phi(p) \to h(p_1) + X(p_2) X(p_3)$, only gets a contribution 
from the squared Riemann tensor. For a graviton polarizations $h_i=+2$ or $-2$, which are described by the tensor 
$\epsilon^{h_i}_{\alpha \beta}= \epsilon^{i}_\alpha \epsilon^{i}_\beta$, with  $i=+1$ or $-1$, the resulting 
amplitude is given as:

\begin{equation}
M^{h_i} = 2 \frac{\kappa \kappa_4}{q^2} [ (q\cdot \epsilon^{i})^2 p^{\rho}_1 p^{\sigma}_1 + \frac{1}{4} q\cdot \epsilon^{i} q\cdot p_1
(p^{\rho}_1 \epsilon^{i\sigma} + p^{\sigma}_1 \epsilon^{i\rho}) + (q\cdot p_1)^2  \epsilon^{i\rho} \epsilon^{i\sigma}] T_{\rho \sigma}
\end{equation}

Then, as an example we consider the decay into a pair of massless scalars ($s$), which amounts to consider the corresponding
Energy-momentum tensor, i.e. $T_{\rho \sigma} =  p_{2\rho}p_{3\sigma} +   p_{3\rho}p_{2\sigma} - \frac{1}{2} p_2\cdot p_3 \eta_{\rho\sigma}$.
The  squared amplitud, summed over polarizations, is given by:
\begin{equation}
\sum_i |M^{h_i}|^2 = 2 (\kappa \kappa_4)^2 (p_1 \cdot p_2)^2[ p_1 \cdot p_2 + \frac{1}{4} p_1 \cdot q]^2
\end{equation}

We can then evaluate the decay width using the formulae for the 3-body phase-space, which can be expressed in terms
of the energy fractions $x_2=2 p^0_2/m$, $x_3=2 p^0_3/m$, such that the differential decay width is given as follows:
 \begin{equation}
\frac{d^2 \Gamma}{dx_2dx_3}     =  \frac{m}{256 \pi^3} \sum_i |M^{h_i}|^2 
\end{equation}
we can then integrate this expression, and get numerical results for the decay width, but we refrain from doing so, since we only wanted to 
find which of the curvature terms in the action contribute to the 3-body  amplitude. Nevertheless, it is interesting to show
that even the 3-body amplitude simplifies this much with the use of  helicity amplitude methods.

\section{Loop effects - graviton self energy}

We have seen that only the squared Riemann tensor contributes to the tree-level amplitude for the decays $\phi \to h h, \, h h^*$, 
and we want to discuss here if this pattern also holds at loop levels. For this we shall discuss the scalar contributions to the graviton self-energy.

From the early work of t'Hooft and Veltman \cite{tHooft:1974toh}, it is known that pure quantum gravity is 
finite at one-loop. This is due to the fact that the possible counter-term is proportional to the Gauss-Bonnet term, 
which is a total derivative in 4D and it can be eliminated from the action. The contribution of scalars to the graviton vacuum polarization has also been studied in detail \cite{Capper:1973bk},  considering the Feynman graphs generated with the
  3-point ($\phi\phi h$) and 4-point vertex ($\phi\phi h h$), from the minimal coupling action.
  These contributions contain a divergence that requires a counter-term involving  $ R, R^2$ and $R_{\mu\nu}R^{\mu\nu}$. 
This term is not contained in the original Lagrangian, thus reflecting the fact that gravity is not renormalizable.
  
The scalar Lagrangian with minimal coupling does not induce a 3-point vertex   $\phi hh$, which can be induced only when
non-minimal couplings are included, i.e. for Higgs multiplets with SSB or scalar singlets. In principle such vertices 
could arise from the squares of the Ricci Scalar, Ricci tensor and Riemann tensor, as well as the second order expansion of the 
Ricci scalar. Thus, we can try to evaluate the contribution from each of these terms to the graviton self-energy, but rather than 
considering all possible diagrams, here we shall only consider the contribution form new type of diagrams, where both an scalar 
and one graviton circulate in the loop (see fig. 1).

\begin{figure}
\centering
\includegraphics[scale=0.3]{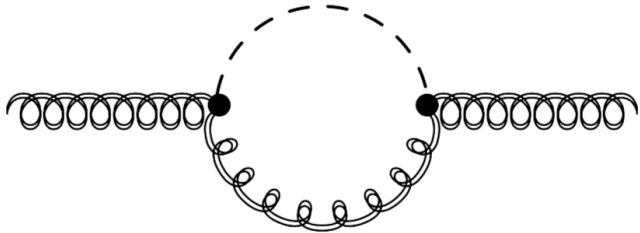}
\caption{Contribution of $\phi hh$ insertion to the graviton self energy}
\label{GG-SlefE}
\end{figure}

Because the loop involves two vertices of the type $h-h^*-\phi$, $V_{\mu\nu\alpha\beta}(k,q)$ and 
$V_{\gamma\eta\rho\sigma} (q, k)$,  with the virtual graviton $h^*$and scalar ($\phi$) circulating the the loop, 
we have that the contributions to the loop coming from the Ricci scalar 
(to second order) and the squares of the Ricci Scalar, Ricci tensor will vanish. Thus the only possible
non-zero contribution of this type will come from the squared Riemann tensor. The amplitude for this diagram
is given by:

\begin{equation}
\Pi_{\mu\nu \rho\sigma} (k_1) = \int \frac{d^4q}{(2\pi)^4} V_{\mu\nu\alpha\beta} (k,q)\frac{iP^{\alpha\beta\gamma\eta}}{q^2}       
                                                                             \frac{i}{(q+k_1)^2-m^2 } V_{\gamma\eta\rho\sigma} (q, k)
\end{equation}
where $m$ is the scalar mass.
However, after contracting with polarization tensor for on-shell gravitons ($k^2=0$), we find that the corresponding amplitude for this diagram also vanishes. Thus, it appears that the physical content of this type of theory is an issue that may be
deeper than we thought.

\section{Conclusions}

In this paper we studied  the coupling of  scalar particles ($\phi$)  with a pair of gravitons ($h$). First, using the 
scaling behavior  of the spinor variables, we derived the general form of the 3-point amplitude  $\phi(p) h(1)h(2)$ 
for the specific  gravitons helicities $h_{i} (1,2) $;  the resulting amplitudes  for the allowed combinations 
$h_i (1,2) = +2, +2$ or  $-2, -2$
turned out to be only of two types, proportional to the spinor products $[kk']^4$ or  $\langle kk' \rangle^4$, which
are of such  simplicity that it motivated us to look for the  simplifications that are suspected to be hidden in the Higgs-graviton action.  
 We  then consider an extended action that  contains the Ricci scalar, Ricci tensor, and Riemann tensor, and expanded these terms
 up to $O(h^2)$ in the graviton field, i.e. $g_{\mu\nu}= \eta_{\mu\nu}+ \kappa h_{\mu\nu}$. We also considered the interactions of a pseudoscalar with gravitons.  We found that the resulting vertices satisfy the KLT relations, and can indeed be written as the square of the YM terms.
 We then constructed the amplitude for decay $\phi \to hh$, and we found that  only the term quadratic in the Riemann tensor
contributes to the decay amplitude; and this remains valid also when we considered  the 3-body decay 
$\phi \to h h^* (\to XX)$, with an off-shell graviton that goes into the final state $XX$. 
Then, in order to find some effect from the quadratic terms in curvature, we considered loop effects.
In particular, we calculated  the contribution to the vertex $h-\phi-h$ to the graviton self-energy, i.e. with Feynman diagrams having one scalar and one graviton in the loop, and we found that this term also vanishes when the external graviton is on-shell.
 
 Although the initial motivation of our work, was to use some phenomenology to search for physical effects of the
interaction of  scalars with gravitons, as we made progress we found that there are extra aspects that one has to 
consider \cite{Alvarez-Gaume:2015rwa}.  It appears that the physical content  of this type of theory that mixes scalars 
with higher-order gravity, is an issue that may be deeper than we thought. 

\bigskip

\bigskip

{\bf{Acknowledgments.}}
Two of us (A.A.-L. and J.L.D.-C.) would like to thank the support of CONACYT and SNI (Mexico).




%

\end{document}